\documentclass[prd,showpacs,nofootinbib,preprint,10 pt]{revtex4}
\usepackage{amsmath,graphicx,color,xcolor,epsfig}
\usepackage{epstopdf}
\usepackage{hyperref}

\begin{document}
\begin{titlepage}

\begin{flushright}
DESY 19-051\\
March 2019
\end{flushright}

\begin{center}
\setlength {\baselineskip}{0.3in} 
{\bf\Large\boldmath 
Interpretation of the Narrow $J/\psi \, p$ Peaks in $\Lambda_b \to J/\psi \, p\, K^-$ Decay \\ 
in the Compact Diquark Model 
}\\[5mm]
\setlength {\baselineskip}{0.2in}
{\large  Ahmed Ali$^{1}$ and Alexander Ya. Parkhomenko$^{2}$}\\[5mm]

$^1$~{\it Deutsches Elektronen-Synchrotron DESY, D-22607 Hamburg, Germany}\\[5mm]

$^2$~{\it Department of Theoretical Physics, P.\,G.~Demidov Yaroslavl State University, \\ 
          Sovietskaya 14, 150003 Yaroslavl, Russia}\\[7mm] 



{\bf Abstract}\\[5mm] 
\end{center}
\setlength{\baselineskip}{0.2in} 
Recently, the LHCb Collaboration have updated their analysis of the resonant $J/\psi\, p$ mass 
spectrum in the decay $\Lambda_b^0 \to J/\psi\, p\, K^-$. In the combined Run~1 and Run~2 LHCb data, 
three peaks are observed, with the former $P_c (4450)^+$ state split into two narrow states, 
$P_c (4440)^+$ and $P_c (4457)^+$, having the masses $M = (4440.3 \pm 1.3^{+4.1}_{-4.7})$~MeV and
$M = (4457.3 \pm 0.6^{+4.1}_{-1.7})$~MeV, and decay widths $\Gamma = (20.6 \pm 4.9^{+8.7}_{-10.1})$~MeV
and $\Gamma = (6.4 \pm 2.0^{+5.7}_{-1.9})$~MeV, respectively. In addition, a third narrow peak, 
$P_c (4312)^+$, having the mass $M = (4311.9 \pm 0.7^{+6.8}_{-0.6})$~MeV and decay width 
$\Gamma = (9.8 \pm 2.7^{+3.7}_{-4.5})$~MeV is also observed. The LHCb analysis is not sensitive 
to broad $J/\psi\, p$ contributions like the former $P_c (4380)^+$, implying that there could be 
more states present in the data. Also the spin-parity, $J^P$, assignments of the states are not 
yet determined. We interpret these resonances in the compact diquark model as hidden-charm 
diquark-diquark-antiquark baryons, having the following spin and angular momentum quantum numbers: 
$P_c (4312)^+ = \{\bar c [cu]_{s=1} [ud]_{s=0}; L_{\mathcal{P}} = 0, J^P = 3/2^- \}$, the $S$-wave 
state, and the other two as $P$-wave states, with 
$P_c (4440)^+ = \{\bar c [cu]_{s=1} [ud]_{s=0}; L_{\mathcal{P}} = 1, J^P = 3/2^+ \}$ and
$P_c (4457)^+ = \{\bar c [cu]_{s=1} [ud]_{s=0}; L_{\mathcal{P}} = 1, J^P = 5/2^+ \}$.
The subscripts denote the spins of the diquarks and $L_{\mathcal{P}} = 0,\,1$ is the orbital 
angular momentum quantum number of the pentaquark. These assignments are in accord with the 
heavy-quark-symmetry selection rules for $\Lambda_b$-baryon decays, in which the spin $S = 0$ 
of the light diquark $[ud]_{s=0}$ is conserved. The masses of observed states can be accommodated
in this framework and the two heaviest states have the positive parities as opposed to the 
molecular-like interpretations. In addition, we predict several more states in the $J/\psi\, p$ 
mass spectrum, and urge the LHCb Collaboration to search for them in their data.
\end{titlepage}

\section{Introduction}
\label{sec:introduction}

In 2015, the LHCb Collaboration reported the first observation of two 
hidden-charm pentaquark states $P_c (4380)^+$ and $P_c (4450)^+$ in the decay 
$\Lambda_b^0 \to J/\psi\, p\, K^-$~\cite{Aaij:2015tga}, having the masses 
$M = (4380 \pm 8 \pm 29)$~MeV and $M = (4449.8 \pm 1.7 \pm 2.5)$~MeV, and 
decay widths $\Gamma = (205 \pm 18 \pm 86)$~MeV and $\Gamma = (39 \pm 5 \pm 19)$~MeV, 
with the preferred spin-parity assignments $J^P = 3/2^-$ and $J^P = 5/2^+$, respectively,
but the reversed spin-parity assignments were also tenable.    
This was followed in a subsequent paper~\cite{Aaij:2016ymb}, in which  evidence 
was presented for the Cabibbo-suppressed decay $\Lambda_b^0 \to J/\psi\, p\, \pi^-$, found 
consistent with the production of the $P_c (4380)^+$ and $P_c (4450)^+$ pentaquark states.
These states have the quark content $(c \bar c u u d)$ and, like their tetraquark 
counterparts~$X$, $Y$, and~$Z$, they lie close in mass to several (charmed meson-baryon) 
thresholds~\cite{Tanabashi:2018oca}. This has led to a number of theoretical proposals 
for their interpretation, which include rescattering-induced kinematical effects~\cite{
Guo:2015umn,Liu:2015fea,Mikhasenko:2015vca,Meissner:2015mza},
open charm-baryon and charm-meson bound states~\cite{
Chen:2015moa,He:2015cea,Roca:2015dva,Chen:2015loa,Xiao:2015fia}, 
and baryocharmonia~\cite{Kubarovsky:2015aaa}. They have also been interpreted as compact 
pentaquark hadrons with the internal structure organised as diquark-diquark-antiquark~\cite{%
Maiani:2015vwa,
Li:2015gta,Mironov:2015ica,Anisovich:2015cia,Ghosh:2015ksa,Wang:2015epa,Wang:2015ava,%
Wang:2015wsa,Ali:2016dkf,Ali:2017ebb}  
or as diquark-triquark~\cite{Lebed:2015tna,Zhu:2015bba}.

Very recently, the LHCb Collaboration has updated their analysis of the 
$\Lambda_b \to J/\psi\, p\, K^-$ decay, making use of 9~times more data based 
on Run~1 and Run~2 than in the Run~1 data alone~\cite{Aaij:2019vzc}.
Thanks to the impressive number (246~K) of $\Lambda_b$-baryon signal events, 
improved $J/\psi\, p$ mass and momentum resolution, and excellent vertexing, 
narrow $J/\psi\, p$ structures have been observed, which were insignificant
in the older Run~1 data~\cite{Aaij:2015tga}. Nominal fits of the data have 
been performed with an incoherent sum of Breit-Wigner amplitudes, which have 
resulted in the observation of three peaks, whose masses, decay widths (with 
95\%~C.L. upper limits) and the ratio~${\cal R}$, defined as 
\begin{equation}
{\cal R} \equiv \frac{{\cal B} (\Lambda_b \to P_c^+\, K^-) \, 
                      {\cal B} (P_c^+ \to J/\psi\, p)}
                     {{\cal B} (\Lambda_b \to J/\psi\, p\, K^-)} ,
\label{eq:R-def}
\end{equation}  
are given in Table~\ref{tab:LHCb-data-2019}. The two narrow resonances, hence-after 
called $P_c (4440)^+$ and $P_c (4457)^+$, replace the older resonance, $P_c (4450)^+$, 
which is now found split, and the third narrow resonance, called $P_c (4312)^+$, is 
a new addition to the pentaquark spectrum. The situation with the broader $P_c (4380)^+$ 
state, observed in the 2015 data, having the large width, $\Gamma = (205 \pm 18 \pm 86)$~MeV,
is not clear, as the current analysis is not sensitive to broad structures. Hence, 
a dedicated analysis of the LHCb data may turn out to have more narrow and broad resonances.

In this work we follow the compact pentaquark interpretation. The basic idea of this 
approach is that highly correlated diquarks play a key role in the physics of multiquark 
states~\cite{Maiani:2004vq,Lipkin:1987sk,Jaffe:2003sg}. 
Since quarks transform as a triplet $\tt 3$ of the color $SU (3)$-group, the diquarks resulting 
from the direct product $\tt 3 \otimes 3 = \bar 3 \oplus 6$, are thus either a color 
anti-triplet~$\tt \bar 3$ or a color sextet~$\tt 6$. Of these only the color~$\tt \bar 3$ 
configuration is kept, as suggested by perturbative arguments. Both spin-1 and spin-0 diquarks 
are, however, allowed. In the case of a diquark $[qq^\prime]$ consisting of two light quarks, 
the spin-0 diquark is believed to be more tightly bound than the spin-1, and this hyperfine 
splitting has implications for the spectroscopy. For the heavy-light diquarks, such as~$[cq]$ 
or~$[bq]$, this splitting is suppressed by $1/m_c$ for the~$[cq]$ or by $1/m_b$ for the~$[bq]$ 
diquark, and hence both spin configurations are treated at par. 
For pentaquarks, the mass spectrum depends upon how the five quarks, i.\,e., the 4~quarks and 
an antiquark, are dynamically structured. A diquark-triquark picture, in which the two observed 
pentaquarks consist of a rapidly separating pair of a color-$\tt \bar 3$ $[cu]$-diquark and 
a color-$\tt 3$ triquark $\bar{\theta} = \bar c [ud]$, has been presented in~\cite{Lebed:2015tna,Zhu:2015bba}. 
A ``Cornell''-type non-relativistic linear-plus-Coulomb potential~\cite{Eichten:1978tg} is used 
to determine the diquark-triquark separation~$R$ and the ensuring phenomenology is worked out. 

\begin{table}
\caption{
Masses, decay widths (with 95\%~C.L. upper limits), and the ratio~$\mathcal{R}$, 
of the three narrow $J/\psi\, p$ resonances observed by the LHCb Collaboration 
in the decay $\Lambda_b \to J/\psi\, p\, K^-$~\cite{Aaij:2019vzc}.}
\label{tab:LHCb-data-2019} 
\begin{center}
\begin{tabular}{ccccc} 
\hline
    State      &         Mass [MeV]             &         Width [MeV]           & (95\% CL) &      $\mathcal{R}\, [\%]$       \\ \hline 
$P_c (4312)^+$ & $4311.9 \pm 0.7^{+6.8}_{-0.6}$ &   $9.8 \pm 2.7^{+3.7}_{-4.5}$ &  ($< 27$) & $0.30 \pm 0.07^{+0.34}_{-0.09}$ \\[1mm] 
$P_c (4440)^+$ & $4440.3 \pm 1.3^{+4.1}_{-4.7}$ & $20.6 \pm 4.9^{+8.7}_{-10.1}$ &  ($< 49$) & $1.11 \pm 0.33^{+0.22}_{-0.10}$ \\[1mm] 
$P_c (4457)^+$ & $4457.3 \pm 0.6^{+4.1}_{-1.7}$ &   $6.4 \pm 2.0^{+5.7}_{-1.9}$ &  ($< 20$) & $0.53 \pm 0.16^{+0.15}_{-0.13}$ \\[1mm] \hline 
\end{tabular}
\end{center} 
\end{table}

In this paper, we employ the diquark-triquark picture shown in Fig.~\ref{ali:fig-pentaquark-model-2}.
Here, the nucleus consists of the doubly-heavy triquark, with the $[c q]$-diquark and 
$\bar c$-antiquark, and a light diquark $[q^\prime q^{\prime\prime}]$, where~$q$, $q^\prime$, 
and~$q^{\prime\prime}$ are~$u$-, $d$-, and $s$-quarks, which is in an orbit for the $P$-wave 
(and higher orbitally-excited) states, as it is easier to excite a light diquark to form 
an orbitally-excited pentaquark. All three constituents are color anti-triplets,~${\tt \bar 3}$, 
making an overall color-singlet~--- the pentaquark. Such a description is closer to the 
doubly-heavy tetraquarks and doubly-heavy baryons, in which the double-heavy diquark may 
be considered as the static color source,    
which has received a lot of theoretical attention lately~\cite{%
Manohar:1992nd,Esposito:2013fma,Luo:2017eub,Karliner:2017qjm,Eichten:2017ffp,Francis:2016hui,%
Bicudo:2017szl,Junnarkar:2017sey,Mehen:2017nrh,Czarnecki:2017vco,Maiani:2019cwl}. 

The resulting pentaquark spectrum in the diquark model is very rich. However, imposing the spin 
conservation in the heavy-quark symmetry limit, which was already advocated in the analysis 
of the older LHCb data~\cite{Ali:2016dkf}, we argue that only that part of the pentaquark spectrum 
is reachable in $\Lambda_b$-decays, in which the pentaquarks have a ``good'' light diquark, i.\,e., 
having the spin $S_{ld} = 0$, in their Fock space. This reduces the number of observable pentaquark 
states greatly. 
There is an overwhelming support of the heavy-quark symmetry 
constraints from the available data on $\Lambda_b$-decays, which are dominated by the 
$\Lambda_b \to \Lambda_c + X$ transitions. Here,~$X$ stands for a large number of leptonic
and hadronic final states, enlisted by Particle Data Group~\cite{Tanabashi:2018oca}. 
The decays $\Lambda_b \to \Sigma_c + X$ are sparse, with PDG listing only two~\cite{Tanabashi:2018oca}: 
$\Lambda_b \to \Sigma_c (2455)^0\, \pi^+\, \pi^-$ and 
$\Lambda_b \to \Sigma_c (2455)^{++}\, \pi^-\, \pi^-$, having the branching fractions 
${\cal B} = (5.7 \pm 2.2) \times 10^{-4}$ and 
${\cal B} = (3.2 \pm 1.6) \times 10^{-4}$, respectively. 
Clearly, there is the $\Sigma_c/\Lambda_c$-suppression in the branching ratios of $\Lambda_b$-decays. 
This also restricts the resonant $J/\psi\, p$ spectrum in the $\Lambda_b \to J/\psi\, p\, K^-$ decay. 

We show that the newly observed spectrum shown in Table~\ref{tab:LHCb-data-2019} 
can be accommodated in the diquark picture, though the spin-parity,~$J^P$, assignment of the observed 
states is tentative, as these quantum numbers are not yet measured. There are yet more states present 
in the $J/\psi\, p$ mass spectrum in the compact diquark model, whose masses and~$J^P$ quantum 
numbers are given here, and we urge the LHCb Collaboration to search for them in their data.

This paper is organised as follows. In Section~\ref{sec:Triquark-Diquark}, we introduce 
the doubly-heavy triquark~--- light diquark model of pentaquarks, and define the state 
vectors having the total angular momentum quantum number~$J$ by 
$\left | S_{hd}, S_t, L_t; S_{ld}, L_{ld}; S, L \right\rangle_J$, 
where~$S_{hd}$, $S_t$, and~$S_{ld}$ are the spins of the heavy diquark, 
doubly-heavy triquark and light diquark, respectively, and~$L_t$ and~$L_{ld}$ 
are the orbital quantum numbers of the triquark and light-diquark, which are combined 
into the total orbital angular momentum~$L$ of the pentaquark. The corresponding set  
of the $S$-wave state vectors (with $L_t = L_{ld} = L = 0$) with the ``good'' 
($S_{ld} = 0$) light diquark is presented in Table~\ref{tab:S-wave-pentaquarks-good-ld}.  
State vectors of the $P$-wave pentaquarks with the ground-state triquark ($L_t = 0$, 
$L_{ld} = L = 1$) and ``good'' light diquark with the spin $S_{ld} = 0$ are given 
in Table~\ref{tab:P-wave-pentaquarks-good-ld}.
In Section~\ref{sec:Effective-Hamiltonian}, we give the effective Hamiltonian 
used to work out the pentaquark mass spectrum. 
In Section~\ref{sec:Mass-Formulae}, the analytical expressions for the matrix elements 
of the effective Hamiltonian are presented taking into account the dominant spin-spin interactions 
in the heavy and light diquarks and in the hidden-charm triquark. For the $P$-wave states, 
additional contributions and mixings due to the orbital and spin-orbit interactions are included. 
In all the cases, we diagonalise the mass matrices, and analytical equations for all the masses 
of the ~$S$- and $P$-wave pentaquarks, which can be reached in $\Lambda_b$-decays, are presented. 
Section~\ref{sec:mass-predictions} contains the values of various input parameters  
and predictions for the pentaquark masses. 
Possible assignments of the newly observed pentaquark states are also discussed here. 
Decay widths of the pentaquarks are briefly discussed in Section~\ref{sec:pentaquark-widths}.
We  conclude in Section~\ref{sec:conclusions}.

\section{Doubly-heavy triquark~--- light diquark model of pentaquarks}
\label{sec:Triquark-Diquark}
In the pentaquark picture considered here, there are two flux tubes, with the first 
stretched between the charmed diquark and charmed antiquark from which the diquark 
is in the color-antitriplet state~$\bar 3$. With the antiquark being also a color 
anti-triplet state~$\bar 3$, their product is decomposed into two irreducible representations, 
$\bar 3 \times \bar 3 = 3 + \bar 6$, from which the color triplet~$3$ is kept. 
This color-triplet triquark makes a color-singlet bound state with the light diquark 
through the second flux tube in much the same way as the quark and diquark bind 
in an ordinary baryon. We modify the effective Hamiltonian for the $S$-wave 
pentaquarks~\cite{Ali:2016dkf,Ali:2017ebb} by keeping the most relevant terms 
for the mass determination, and then extend it for the $P$-states by including 
the orbital and spin-orbit interactions between the hidden-charm triquark and light diquark. 
Tensor interactions affect the $P$-wave pentaquark spectrum in which light diquarks 
have the spin $S_{ld} = 1$.

The effective Hamiltonian for the ground-state pentaquarks is described in terms 
of two constituent masses of the heavy diquark~$m_{[cq]} \equiv m_{hd}$ 
and the light diquark~$m_{[q^{\prime}q^{\prime\prime}]} \equiv m_{ld}$, where~$q$, 
$q^\prime$, and~$q^{\prime\prime}$ are light~$u$-, $d$- and $s$-quarks, the spin-spin 
interactions between the quarks in each diquark shell, and spin-spin interactions 
between the diquarks. To these are added the constituent mass~$m_c$ of the charmed 
antiquark and its spin-spin interactions with each of the diquarks.

In the case of the orbitally excited pentaquarks, the orbital angular momentum~$L$ 
of the pentaquark is the sum of two terms,~$L_t$, arising from the triquark system 
consisting of the heavy diquark and the charmed antiquark, and $L_{ld}$, which 
determines the relative motion of the light diquark around the doubly-heavy triquark. 
The total orbital angular momentum~$L$ of the pentaquark is then obtained with the help 
of the momentum sum rules from quantum mechanics, i.\,e., it takes a value from the range 
$L = |L_t - L_{ld}|, \ldots, L_t + L_{ld}$. The orbital angular momentum~$L$ is combined 
with the spin~$S$ to get the total angular momentum~$J$ of the pentaquark using
the $L-S$ coupling scheme. 
With the quantum numbers introduced above, we specify the complete orthogonal 
set of basis vectors for the hidden-charm pentaquark states. As already stated, 
we define the state vectors having a total angular momentum quantum number~$J$ 
by $\left | S_{hd}, S_t, L_t; S_{ld}, L_{ld}; S, L \right\rangle_J$.

\begin{figure}
\centerline{\includegraphics[scale=0.4]{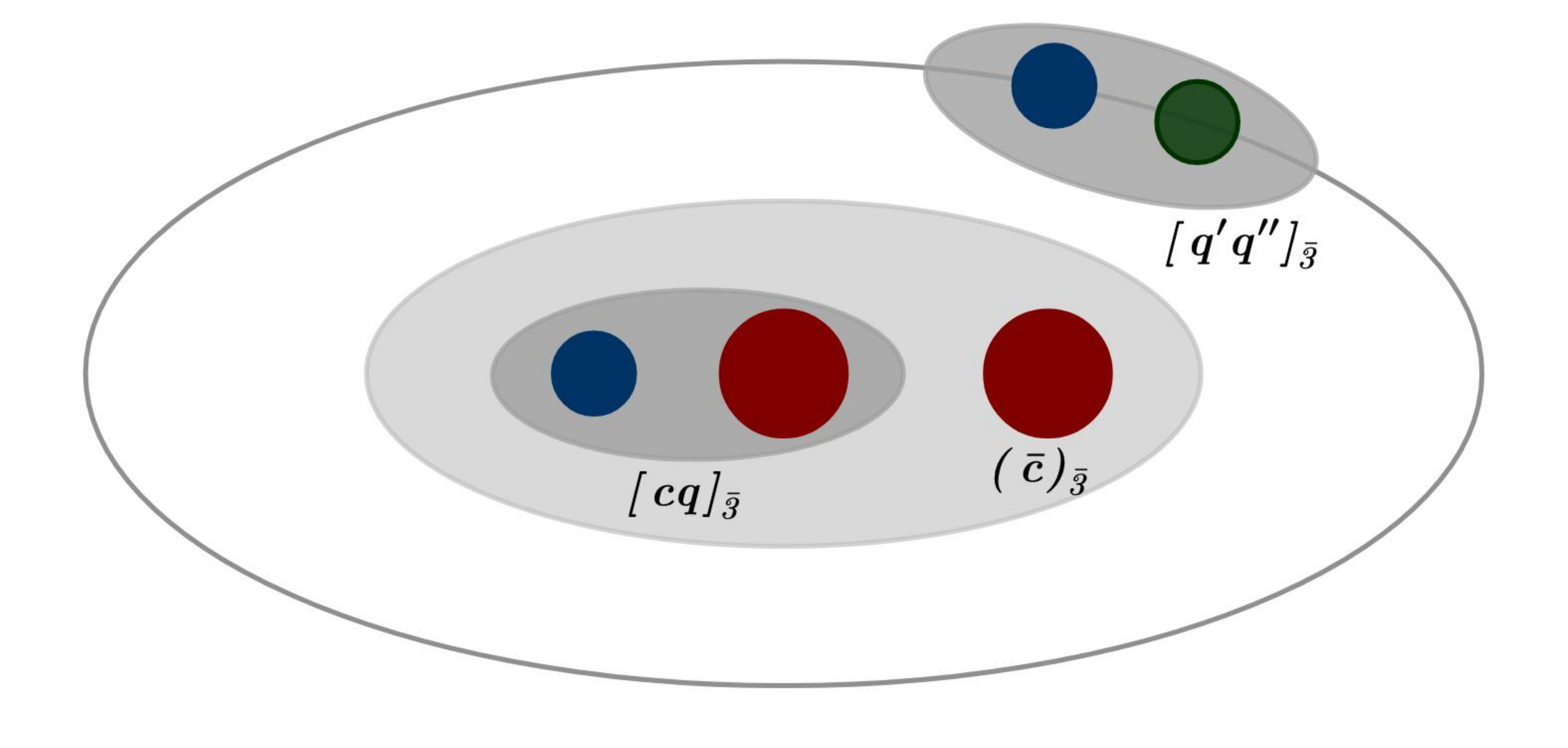}}
\vspace*{-4mm}
\caption{
A picture of pentaquarks in the diquark model involving the heavy diquark 
$[c q]_{\bar 3}$ and charmed antiquark $(\bar c)_{\bar 3}$, which form 
the triquark system, which combines with the light diquark $[q^\prime q^{\prime\prime}]_{\bar 3}$ 
to make a color singlet pentaquark. The subscripts indicate that all three 
constituents are color anti-triplets;~$q$, $q^\prime$, and~$q^{\prime\prime}$ are light quarks 
each of which can be~$u$-, $d$-, or $s$-quark.
}
\label{ali:fig-pentaquark-model-2}
\end{figure}

For the orbitally excited states, one needs to specify which part of the pentaquark 
wave function is excited. In the triquark-diquark template used here, and shown in 
Fig.~\ref{ali:fig-pentaquark-model-2}, the heavy triquark state consists of the charmed 
diquark and charmed antiquark. Since both are heavy, the triquark is an (almost) static 
system. Hence, $L_t = 0$ is the most probable quantum state of the triquark. Thus, the 
orbital excitation is generated by the light diquark (i.\,e., $L = L_{ld}$). With this, 
we list the lowest-lying orbitally excited states ($L = 1$) with the ``good'' ($S_{ld} = 0$) 
light diquark in Table~\ref{tab:P-wave-pentaquarks-good-ld}.

\begin{table}[tb] 
\begin{minipage}[ct]{0.45\textwidth}
\caption{
Spin-parity~$J^P$ and state vectors of the $S$-wave pentaquarks 
containing the ``good'' light diquark with the spin $S_{ld} = 0$.
The horizontal line demarcates the spin~$S_{hd}$ of the heavy diquark.
}
\label{tab:S-wave-pentaquarks-good-ld}
\begin{tabular}{l|l} 
\hline\hline 
  $J^P$ & $\left | S_{hd}, S_t, L_t; S_{ld}, L_{ld};   S, L \right\rangle_J$    \\ 
\hline\hline
$1/2^-$ &      $\left | 0, 1/2,   0;   0,   0; 1/2, 0 \right\rangle_{1/2}$ \\ \hline 
$1/2^-$ &      $\left | 1, 1/2,   0;   0,   0; 1/2, 0 \right\rangle_{1/2}$ \\
$3/2^-$ &      $\left | 1, 3/2,   0;   0,   0; 3/2, 0 \right\rangle_{3/2}$ \\  
\hline\hline 
\end{tabular}
\end{minipage}
\qquad\qquad
\begin{minipage}[ct]{0.45\textwidth}
\caption{
Spin-parity~$J^P$ and state vectors of the $P$-wave pentaquarks with the ground-state 
triquark ($L_t = 0$) and ``good'' light diquark with the spin $S_{ld} = 0$.
The horizontal line demarcates the spin~$S_{hd}$ of the heavy diquark.
}
\label{tab:P-wave-pentaquarks-good-ld}
\begin{tabular}{l|l} 
\hline\hline 
  $J^P$ & $\left | S_{hd}, S_t, L_t; S_{ld}, L_{ld};  S, L \right\rangle_J$    \\ 
\hline\hline 
$1/2^+$ &      $\left | 0, 1/2,   0;   0,   1; 1/2, 1 \right\rangle_{1/2}$ \\ 
$3/2^+$ &      $\left | 0, 1/2,   0;   0,   1; 1/2, 1 \right\rangle_{3/2}$ \\ \hline 
$1/2^+$ &      $\left | 1, 1/2,   0;   0,   1; 1/2, 1 \right\rangle_{1/2}$ \\
$3/2^+$ &      $\left | 1, 1/2,   0;   0,   1; 1/2, 1 \right\rangle_{3/2}$ \\
$1/2^+$ &      $\left | 1, 3/2,   0;   0,   1; 3/2, 1 \right\rangle_{1/2}$ \\
$3/2^+$ &      $\left | 1, 3/2,   0;   0,   1; 3/2, 1 \right\rangle_{3/2}$ \\
$5/2^+$ &      $\left | 1, 3/2,   0;   0,   1; 3/2, 1 \right\rangle_{5/2}$ \\ 
\hline\hline 
\end{tabular}
\end{minipage}
\end{table}

\section{Effective Hamiltonian for Pentaquark Spectrum}
\label{sec:Effective-Hamiltonian}
We calculate the mass spectrum of pentaquarks under the assumption that their 
underlying structure is given by $\bar c$, $[cq]$, and $[q^\prime q^{\prime \prime}]$, 
with~$q$, $q^\prime$, and~$q^{\prime \prime}$ being any of the light $u$-, $d$-, 
and $s$-quarks.  For this, we extend the effective Hamiltonian proposed for 
the tetraquark spectroscopy~\cite{Maiani:2014aja}. The effective Hamiltonian 
for the $S$-wave pentaquark mass spectrum can be written as follows:  
\begin{equation}
H^{(L = 0)} = H_t + H_{ld} . 
\label{eq:Hamiltonian-S-wave} 
\end{equation}
The first term in the Hamiltonian~(\ref{eq:Hamiltonian-S-wave})  
is related with the colored triquark: 
\begin{equation}
H_t = m_c + m_{hd} + 
2 \, (\mathcal{K}_{cq})_{\bar 3} \, (\mathbf{S}_c \cdot \mathbf{S}_q) + 
2 \, \mathcal{K}_{\bar c q} \left ( \mathbf{S}_{\bar c} \cdot \mathbf{S}_q \right ) +    
2 \, \mathcal{K}_{\bar c c} \left ( \mathbf{S}_{\bar c} \cdot \mathbf{S}_c \right ) ,
\label{eq:H-triquark}
\end{equation}
where~$m_c$ and~$m_{hd}$ are the constituent masses of the charmed antiquark and charmed 
diquark, respectively. The last three terms describe the spin-spin interactions in the 
charmed diquark and between the diquark constituents and the charmed antiquark, quantified by
 the couplings~$(\mathcal{K}_{cq})_{\bar 3}$, $\mathcal{K}_{\bar c q}$, 
and~$\mathcal{K}_{\bar c c}$.

The second term~$H_{ld}$ in the Hamiltonian~(\ref{eq:Hamiltonian-S-wave}) contains 
the operators responsible for the spin-spin interaction in the light diquark and 
its interaction with the triquark: 
\begin{equation}
H_{ld} = m_{ld} + 
2 \, (\mathcal{K}_{q^\prime q^{\prime \prime}})_{\bar 3} \, 
(\mathbf{S}_{q^\prime} \cdot \mathbf{S}_{q^{\prime\prime}}) + 
H_{SS}^{t-ld} , 
\label{eq:H-tri-diquark-interaction}
\end{equation}
where $m_{ld}$ is the constituent mass of the light diquark, consisting 
of the light quarks~$q^\prime$ and~$q^{\prime\prime}$, and the contribution 
of the spin-spin interaction in this diquark to the pentaquark mass is 
determined by the coupling~$(\mathcal{K}_{q^\prime q^{\prime\prime}})_{\bar 3}$. 
The last term in~$H_{ld}$ is responsible for all possible spin-spin interactions 
between the constituents of the light diquark and heavy triquark: 
\begin{eqnarray}
H_{SS}^{t-ld} & = & 
2 \, (\tilde{\mathcal{K}}_{c q^\prime})_{\bar 3} \, (\mathbf{S}_c \cdot \mathbf{S}_{q^\prime}) + 
2 \, (\tilde{\mathcal{K}}_{q q^\prime})_{\bar 3} \, (\mathbf{S}_q \cdot \mathbf{S}_{q^\prime}) + 
2 \, \tilde{\mathcal{K}}_{\bar c q^\prime} \, (\mathbf{S}_{\bar c} \cdot \mathbf{S}_{q^\prime})  
\nonumber \\ 
& + & 
2 \, (\tilde{\mathcal{K}}_{c q^{\prime\prime}})_{\bar 3} \, (\mathbf{S}_c \cdot \mathbf{S}_{q^{\prime\prime}}) + 
2 \, (\tilde{\mathcal{K}}_{q q^{\prime\prime}})_{\bar 3} \, (\mathbf{S}_q \cdot \mathbf{S}_{q^{\prime\prime}}) + 
2 \, \tilde{\mathcal{K}}_{\bar c q^{\prime\prime}} \, (\mathbf{S}_{\bar c} \cdot \mathbf{S}_{q^{\prime\prime}}) . 
\label{eq:H-SS-t-d}
\end{eqnarray}
Here, the coefficients with the tilde differ from the ones introduced earlier in Eq.~(\ref{eq:H-triquark}), as they are
the couplings of the spin-spin interactions inside the compact objects, 
like diquarks and triquark,  while the ones in Eq.~(\ref{eq:H-SS-t-d}) denote the couplings between the 
constituents of the two objects, the heavy triquark and light diquark, and are anticipated to be strongly 
suppressed. This then accounts for all possible spin-spin interactions and completes 
the content of the effective Hamiltonian~(\ref{eq:Hamiltonian-S-wave}) for the masses 
of the ground-state pentaquarks.

The general form of the effective Hamiltonian for the orbitally-excited pentaquark 
mass spectrum can be written as follows:  
\begin{equation}
H = H^{(L = 0)} + H_L + H_T . 
\label{eq:Hamiltonian-general} 
\end{equation}
In addition to the spin-spin interactions introduced for the ground-state pentaquarks
described above, the extended effective Hamiltonian~(\ref{eq:Hamiltonian-general})  
includes terms explicitly dependent on the internal orbital angular momentum~$L_t$ 
of the hidden-charm triquark and orbital momentum~$L_{ld}$ of the light diquark 
relative to the triquark system. The corresponding Hamiltonian is called~$H_L$ 
in~(\ref{eq:Hamiltonian-general}). The  tensor interactions 
are subsumed in~$H_T$ in~(\ref{eq:Hamiltonian-general}). 
They are relevant only for the pentaquarks in which the light-diquarks have the spin 
$S_{ld} = 1$. Since they are not anticipated to be produced in the $\Lambda_b$-baryons 
decays, due to the heavy-quark symmetry constraints, we set $H_T = 0$ in this paper. 
They have to be included in the extended pentaquark systems, detailed in  
a forthcoming paper~\cite{Ali:2019xyz}.
As already stated, we assume that the triquark state is an $S$-wave (i.\,e., $L_t = 0$). 
Thus, the total angular momentum of the pentaquark is determined 
by the orbital excitation of the light diquark, $L = L_{ld}$.
The terms in~$H_L$ which contain the orbital angular momentum 
operator~$\mathbf{L}$ are as follows: 
\begin{equation}
H_L = 2 A_t    \left ( \mathbf{S}_t    \cdot \mathbf{L} \right ) +  
      2 A_{ld} \left ( \mathbf{S}_{ld} \cdot \mathbf{L} \right ) + 
      \frac{1}{2} \, B \, \mathbf{L}^2 ,  
\label{eq:Hamiltonian-L-pentaquark} 
\end{equation}
where~$A_t$, $A_{ld}$, and~$B$ parametrise the strengths of the triquark 
spin-orbit, light-diquark spin-orbit and orbital momentum couplings, respectively.

\section{Mass Formulas for Pentaquark Spectrum}
\label{sec:Mass-Formulae} 

\subsection{$S$-wave pentaquarks}
\label{sec:Mass-Formulae-S-wave} 

With the basis vectors of the pentaquark states chosen, one can derive analytical 
expressions for calculating the pentaquark spectrum~\cite{Ali:2019xyz}. They are the matrix 
elements of the effective Hamiltonian presented in Sec.~\ref{sec:Effective-Hamiltonian}. 
Note that this is simpler for the Model~II by Maiani {\it et al.}~\cite{Maiani:2014aja}, 
but becomes more involved in the Model~I~\cite{Maiani:2004vq}, where additional 
interactions~(\ref{eq:H-SS-t-d}) between the spins of (anti)quarks in compact shells 
are included. As the later couplings are suppressed in comparison with the spin-spin 
ones inside the shells, we neglect the contributions from the term~(\ref{eq:H-SS-t-d}) 
and restrict ourselves with the Model~II~\cite{Maiani:2014aja}.

The universal contribution entering all the pentaquark states is defined as~$M_0$, 
which is the sum of the constituent masses of the heavy and light diquarks and 
charmed antiquark: 
\begin{equation} 
M_0 \equiv m_{hd} + m_{ld} + m_c . 
\label{eq:M0-def}
\end{equation} 
Apart from this, there are two terms in the effective Hamiltonian explicitly 
related with the spins of the diquarks [see Eqs.~(\ref{eq:H-triquark}) 
and~(\ref{eq:H-tri-diquark-interaction})]: 
\begin{eqnarray} 
&& 
{}_J \langle S_{hd}, S_t, L_t; S_{ld}, L_{ld}; S, L | \, 
2 \, (\mathcal{K}_{cq})_{\bar 3} \, (\mathbf{S}_c \cdot \mathbf{S}_q) \,    
| S_{hd}, S_t, L_t; S_{ld}, L_{ld}; S, L \rangle_J 
\label{eq:hd-spin-contribution} \\ 
&& \hspace{17mm}
= (\mathcal{K}_{cq})_{\bar 3} 
\left [ S_{hd} \left ( S_{hd} + 1 \right ) - \frac{3}{2} \right ] = 
\frac{1}{2} \, (\mathcal{K}_{cq})_{\bar 3} \times 
\left \{ 
\begin{array}{cc} 
- 3, & S_{hd} = 0, \\ 
  1, & S_{hd} = 1,   
\end{array}
\right. 
\nonumber   
\end{eqnarray}
\begin{eqnarray} 
&& 
{}_J \langle S_{hd}, S_t, L_t; S_{ld}, L_{ld}; S, L | \, 
2 \, (\mathcal{K}_{q^\prime q^{\prime\prime}})_{\bar 3} \, 
(\mathbf{S}_{q^\prime} \cdot \mathbf{S}_{q^{\prime\prime}}) \,   
| S_{hd}, S_t, L_t; S_{ld}, L_{ld}; S, L \rangle_J 
\label{eq:ld-spin-contribution} \\ 
&& \hspace{17mm} = 
(\mathcal{K}_{q^\prime q^{\prime\prime}})_{\bar 3} 
\left [ S_{ld} \left ( S_{ld} + 1 \right ) - \frac{3}{2} \right ] = 
\frac{1}{2} \, (\mathcal{K}_{q^\prime q^{\prime\prime}})_{\bar 3} \times 
\left \{ 
\begin{array}{cc} 
- 3, & S_{ld} = 0, \\ 
  1, & S_{ld} = 1.    
\end{array}
\right.  
\nonumber 
\end{eqnarray}

The remaining terms in Eq.~(\ref{eq:H-triquark}) are responsible for the contributions  
of the spin-spin interactions between the charmed antiquark and the quarks inside 
the heavy diquark, which together form the triquark state. To calculate their impact, 
it is convenient to use the Wigner $6j$-symbols, which allows us to describe a recoupling 
between three spins inside the triquark. The details of calculations are given
in~\cite{Ali:2019xyz} and we present the results here.

From the three $S$-wave states presented in Table~\ref{tab:S-wave-pentaquarks-good-ld}, 
the two states with $J^P = 1/2^-$ mix due to the spin-spin interaction of the charmed 
antiquark and the heavy diquark, and the third one with $J^P = 3/2^-$ remains unmixed.

The two states with $J^P = 1/2^-$, after sandwiching the effective Hamiltonian, yield 
the following $(2 \times 2)$ mass matrix: 
\begin{equation}  
M^{S0}_{J = 1/2} = M_0 - 
\frac{1}{2} \, (\mathcal{K}_{c q})_{\bar 3} - 
\frac{3}{2} \, (\mathcal{K}_{q^\prime q^{\prime\prime}})_{\bar 3} - 
(\mathcal{K}_{c q})_{\bar 3} 
\left ( 
\begin{array}{rr} 
 1 &  0 \\
 0 & -1
\end{array}
\right ) 
+ \frac{1}{2} \left ( \mathcal{K}_{\bar c q} + \mathcal{K}_{\bar c c} \right )  
\left ( 
\begin{array}{rr} 
       0 & \sqrt 3 \\
 \sqrt 3 &      -2
\end{array}
\right ) .  
\label{eq:MM-S0-1/2} 
\end{equation}
Diagonalising this matrix yields the masses of these two states:
\begin{eqnarray} 
m^{S0}_1 = M_0 - 
\frac{1}{4} \left ( \mathcal{K}_{\bar c q} + \mathcal{K}_{\bar c c} \right )  
\left [ 2 + r_{hd} + 3 r_{ld} + 2 \sqrt{3 + (1 - r_{hd})^2} \right ] ,  
\label{eq:mass-S0-1} \\ 
m^{S0}_2 = M_0 - 
\frac{1}{4} \left ( \mathcal{K}_{\bar c q} + \mathcal{K}_{\bar c c} \right )  
\left [ 2 + r_{hd} + 3 r_{ld} - 2 \sqrt{3 + (1 - r_{hd})^2} \right ] ,  
\label{eq:mass-S0-2} 
\end{eqnarray}
where the superscript~$S0$ denotes the $S$-wave pentaquark with the ``good'' light diquark, 
$S_{ld} = 0$, and the two ratios~$r_{hd}$ and~$r_{ld}$ of the couplings are defined as:  
\begin{equation}
r_{hd} \equiv \frac{2 (\mathcal{K}_{c q})_{\bar 3}}{
               \mathcal{K}_{\bar c q} + \mathcal{K}_{\bar c c}} , 
\qquad  
r_{ld} \equiv \frac{2 (\mathcal{K}_{q^\prime q^{\prime\prime}})_{\bar 3}}{ 
               \mathcal{K}_{\bar c q} + \mathcal{K}_{\bar c c}} . 
\label{eq:rhd-rld-def}
\end{equation} 

For the state with $J^P = 3/2^-$, the mass~$m^{S0}_3$ is the average 
of the effective Hamiltonian over this state: 
\begin{equation}  
m^{S0}_3 = M_0 + 
\frac{1}{4} \left ( \mathcal{K}_{\bar c q} + \mathcal{K}_{\bar c c} \right ) 
\left ( 2 + r_{hd} - 3 r_{ld} \right ) .  
\label{eq:mass-S0-3} 
\end{equation}

The case of $S$-wave pentaquark states with the ``bad'' light diquark 
is worked out in detail elsewhere~\cite{Ali:2019xyz}.

\subsection{$P$-wave pentaquarks}
\label{sec:Mass-Formulae-P-wave} 
As  discussed earlier in Sec.~\ref{sec:Effective-Hamiltonian}, one needs to include the
terms dependent on the internal angular momenta of the pentaquark and, in particular, 
the term~(\ref{eq:Hamiltonian-L-pentaquark}) from the pentaquark effective Hamiltonian.

From the seven $P$-wave pentaquark states with the ``good'' light diquark shown in 
Table~\ref{tab:P-wave-pentaquarks-good-ld}, two states 
with $J^P = 1/2^+$ and the other two with $J^P = 3/2^+$, both pairs having the triquark 
spin $S_t = 1/2$, mix due to the spin-spin interaction of the charmed antiquark and heavy 
diquark. The other three states with the triquark spin $S_t = 3/2$ remain unmixed due to 
this interaction but can mix through the spin-spin interactions between the (anti)quarks 
entering two separated shells~--- the heavy triquark and light diquark, as also discussed 
earlier for the $S$-wave states. As mentioned earlier, such types of spin-spin interactions 
are suppressed and neglected in this analysis.
From the angular-momentum dependent term~$H_L$ 
of the effective Hamiltonian, one obtains the spin-orbit and orbital contributions 
to matrix elements.

The pair of states with $J^P = 1/2^+$, after sandwiching the effective 
Hamiltonian, yields the following $(2 \times 2)$ mass matrix: 
\begin{equation}  
M^{P0}_{J = 1/2} = M_0 - 
\frac{1}{2} \, (\mathcal{K}_{c q})_{\bar 3} - 
\frac{3}{2} \, (\mathcal{K}_{q^\prime q^{\prime\prime}})_{\bar 3} - 
(\mathcal{K}_{c q})_{\bar 3} 
\left ( 
\begin{array}{rr} 
 1 &  0 \\
 0 & -1
\end{array}
\right ) 
+ \frac{1}{2} \left ( \mathcal{K}_{\bar c q} + \mathcal{K}_{\bar c c} \right )  
\left ( 
\begin{array}{rr} 
       0 & \sqrt 3 \\
 \sqrt 3 &      -2
\end{array}
\right ) + B - 2 A_t ,  
\label{eq:MM-P0-1/2} 
\end{equation}
while the masses of the pair with $J^P = 3/2^+$ are determined by the matrix:  
\begin{equation}  
M^{P0}_{J = 3/2} = M_0 - 
\frac{1}{2} \, (\mathcal{K}_{c q})_{\bar 3} - 
\frac{3}{2} \, (\mathcal{K}_{q^\prime q^{\prime\prime}})_{\bar 3} - 
(\mathcal{K}_{c q})_{\bar 3} 
\left ( 
\begin{array}{rr} 
 1 &  0 \\
 0 & -1
\end{array}
\right ) 
+ \frac{1}{2} \left ( \mathcal{K}_{\bar c q} + \mathcal{K}_{\bar c c} \right )  
\left ( 
\begin{array}{rr} 
       0 & \sqrt 3 \\
 \sqrt 3 &      -2
\end{array}
\right ) + B + A_t .   
\label{eq:MM-P0-3/2} 
\end{equation}
Diagonalising these matrices, we get the masses: 
\begin{eqnarray} 
&& 
m^{P0}_1 = M_0 - 
\frac{1}{4} \left ( \mathcal{K}_{\bar c q} + \mathcal{K}_{\bar c c} \right )  
\left [ 2 + r_{hd} + 3 r_{ld} + 2 \sqrt{3 + (1 - r_{hd})^2} \right ] + B - 2 A_t ,  
\label{eq:mass-P0-1} \\ 
&& 
m^{P0}_2 = m^{P0}_1 +  
\left ( \mathcal{K}_{\bar c q} + \mathcal{K}_{\bar c c} \right ) \sqrt{3 + (1 - r_{hd})^2} ,  
\label{eq:mass-P0-2} \\ 
&& 
m^{P0}_{3,4} = m^{P0}_{1,2} + 3 A_t ,  
\label{eq:mass-P0-3-4}  
\end{eqnarray}
where~$r_{hd}$ and $r_{ld}$ are defined in Eq.~(\ref{eq:rhd-rld-def}) and 
the superscript~$P0$ means the $P$-wave pentaquark with the ``good'' light 
diquark, $S_{ld} = 0$.

For the three states, enumerated as the fifth, sixth and seventh according to the 
entries in Table~\ref{tab:P-wave-pentaquarks-good-ld}, the masses~$m^{P0}_{5,6,7}$ 
are simply the average values of the effective Hamiltonian over these states: 
\begin{eqnarray}  
&& 
m^{P0}_5 = M_0 + 
\frac{1}{4} \left ( \mathcal{K}_{\bar c q} + \mathcal{K}_{\bar c c} \right ) 
\left ( 2 + r_{hd} - 3 r_{ld} \right ) 
+ B - 5 A_t , 
\label{eq:mass-P0-5} \\ 
&& 
m^{P0}_6 = m^{P0}_5 + 3 A_t , 
\qquad  
m^{P0}_7 = m^{P0}_5 + 8 A_t .  
\nonumber 
\end{eqnarray}
These states are non-degenerate due to the triquark spin-orbit interaction, i.\,e., $A_t \neq 0$. 
Hence, the mass gap between them is a measure of the coupling~$A_t$, and the following 
relations hold: 
\begin{equation} 
m^{P0}_6 - m^{P0}_5 = 
\frac{3}{8} \left ( m^{P0}_7 - m^{P0}_5 \right ) =  
\frac{3}{5} \left ( m^{P0}_7 - m^{P0}_6 \right ).
\label{eq:m-P0-relation}
\end{equation}
If more states are experimentally observed and their quantum numbers determined, 
the validity of these relations will be a strong argument in favour of the compact 
diquark model.

\section{Hidden-charm Pentaquark Mass Predictions}                      
\label{sec:mass-predictions}

Working within the constituent quark-diquark model~\cite{Maiani:2014aja}, input parameters 
are the masses of the constituents, charmed quark and diquarks, spin-spin couplings, 
and other parameters related with the orbital or radial excitations. 
For the constituent quark masses, there are two possibilities: Either extract 
them from the masses of the known mesons or from baryons~\cite{Ali:2019}.  
The charmed quark mass  is estimated as $m_c^m = 1667$~MeV 
from the $D$-mesons spectrum, and $m_c^b = 1710$~MeV from the charmed baryon masses, 
yielding a difference of~$43$~MeV~\cite{Ali:2019}. In particular, with the~$m_c^b$ value 
as an input, predictions for the charmed baryon masses were obtained in~\cite{Karliner:2014gca},
which are remarkably accurate and differ from the experimentally measured 
masses~\cite{Tanabashi:2018oca} by about $10$~MeV. We use~$m_c^b$ in the numerical analysis.  

In contrast to quarks, which have the fixed spin $S_q = 1/2$, for diquarks being 
composite objects, there are two possible spin configurations from which the 
antisymmetric one corresponding to the diquark spin $S = 0$ is energetically more 
favourable. Both configurations are allowed if the flavors of the quarks are different. 
As we consider the pentaquarks from the $\Lambda_b$-decay in the heavy-quark symmetry 
limit, we need to specify the mass of the light $[ud]$-diquark, for which we use
the value $m_{ld} = m_{[ud]} = 576$~MeV~\cite{Karliner:2018bms}.  
The charmed diquarks mass $m_{hd} = m_{[c q]} = 1976$~MeV is borrowed from~\cite{Ali:2019}. 
To estimate the errors on the diquark-masses, one should compare the baryon masses predicted 
within this (constituent quark-diquark) model and the  measured ones. Based on the analysis 
presented in~\cite{Karliner:2014gca}, we infer that the computed masses for the charmed baryons
are accurate to within $15$~MeV. This value may be taken as indicative of the uncertainty 
in diquark masses.      

In addition, we need the spin-spin couplings,~$\mathcal{K}_{\bar Q Q^\prime}$ 
and~$(\mathcal{K}_{Q Q^\prime})_{\bar 3}$, which are extracted from the mass spectra 
of mesons and baryons, respectively. The values of the spin-spin couplings used are:  
$(\mathcal{K}_{cq})_{\bar 3} = 67$~MeV, $(\mathcal{K}_{q q^\prime})_{\bar 3} = 98$~MeV, 
$\mathcal{K}_{\bar c q} = 70$~MeV, and $\mathcal{K}_{\bar c c} = 113$~MeV. They are taken 
from~\cite{Ali:2019}, except for $\mathcal{K}_{\bar c c}$ which is from~\cite{Karliner:2014gca}. 
For numerical analysis, one needs to show uncertainces in these couplings and we assign 
it to be~10\% of each value, the same as in the $\Omega_c^*$-baryons~\cite{Ali:2017wsf}.  
The pentaquark masses involve the ratios of the couplings~(\ref{eq:rhd-rld-def}) 
which for the input values are evaluated as: $r_{hd} = 0.73$ and~$r_{ld} = 1.07$.  
In Model~II~\cite{Maiani:2014aja}, these ratios are of $O (1)$, which reflects 
the dominance of the spin-spin interaction inside the diquarks and triquark over 
other possible spin-spin interactions. These arguments give quantitative support 
in favour of the Model~II~\cite{Maiani:2014aja}.
   
In estimating the $P$-wave pentaquark mass spectrum, the values of the couplings 
in the orbital angular momentum term~$B$ and the spin-orbit coupling~$A_t$ are required. 
We discuss first the two heaviest pentaquarks $P_c (4440)^+$ and $P_c (4457)^+$. 
As they replace the former narrow $P_c (4450)^+$ state with the preferred spin-parity 
$J^P = 5/2^+$, we tentatively assign this spin-parity to one of the observed 
two states, $P_c (4457)^+$. In this case, the lighter partner, $P_c (4440)^+$, most 
probably has the spin-parity $J^P = 3/2^+$. In the Model~II, used here for the pentaquark 
spectrum, their mass splitting is related to the spin-orbit coupling~$A_t$ of the triquark: 
\begin{equation} 
M [P_c (4457)^+] - M [P_c (4440)^+] = \left ( 17^{+6.4}_{-4.5} \right )~\mathrm{MeV} = 5 A_t , 
\label{eq:Pc-76-mass-diff}
\end{equation}  
where the error is obtained by adding the experimental errors on the masses 
in quadrature. From~(\ref{eq:Pc-76-mass-diff}), the value of the coefficient~$A_t$ 
follows immediately: 
\begin{equation} 
A_t = \left ( 3.4^{+1.3}_{-0.9} \right )~\mathrm{MeV} .  
\label{eq:At-experiment}
\end{equation}  
It is not surprising that it is numerically small, as the doubly-heavy 
triquark is almost static. Of course, as there are more states present in the
theoretical spectrum than the number of observed pentaquarks, there are also other
assignments possible, but we find them unrealistic and hence not discussed here.

The third narrow state, $P_c (4312)^+$, can have several assignments. 
Identifying it with the $J^P = 3/2^-$ state, one can work out the mass difference 
between this state and the heavier pentaquarks, $P_c (4440)^+$ and $P_c (4457)^+$. 
This is determined by the orbital~$B$ and the triquark spin-orbit~$A_t$ couplings. 
The strength of the later coupling is already known from the $P_c (4440)^+$ 
and~$P_c (4457)^+$ mass splitting~(\ref{eq:At-experiment}), and the mass difference, 
say, between $P_c (4312)^+$ and $P_c (4457)^+$, allows us to read off the strength 
of the orbital interaction: 
\begin{equation}
M [P_c (4457)^+] - M [P_c (4312)^+] = \left ( 145.4^{+4.2}_{-7.1} \right )~\mathrm{MeV} = B + 3 A_t . 
\label{eq:Pc-73-mass-diff}
\end{equation}  
With $A_t$ from~(\ref{eq:At-experiment}), we get:   
\begin{equation} 
B = \left ( 135.2^{+5.0}_{-8.1} \right )~\mathrm{MeV} .  
\label{eq:B-splitting-exp}
\end{equation}  
This is too small in comparison with the strengths of the orbital excitations in other 
hadrons~\cite{Ali:2017wsf}, and, in particular, with $B (\Omega_c) = 325$~MeV obtained 
from $\Omega_c^*$-baryons. Moreover, the theoretically predicted masses of the $P_c (4440)^+$ 
and $P_c (4457)^+$ states with the value of~$B$ from~(\ref{eq:B-splitting-exp}) are found 
to be $\sim 70$~MeV below the experimental values.   
Alternatively, assuming that the spin-spin couplings and constituent masses are known, 
the strength of~$B$ can also be determined from the masses of $P_c (4440)^+$ 
and $P_c (4457)^+$ only, as follows:  
\begin{equation}
B = \frac{1}{5} \left \{ 3 M [P_c (4440)^+] + 2 M [P_c (4457)^+] \right \} 
- M_0 - \frac{1}{4} \left ( \mathcal{K}_{\bar c q} + \mathcal{K}_{\bar c c} \right ) 
\left ( 2 + r_{hd} - 3 r_{ld} \right ) . 
\label{eq:Pc-76-mass-sum}
\end{equation}  
With the values of the other parameters already assigned, the value $B = 207$~MeV  
reproduces the masses of the observed $P_c(4440)$ and $P_c(4457)$ states,
as shown in Table~\ref{tab:masses-predictions-cbar-cq-qq} as the last two entries 
inserted into the solid boxes. This value of~$B$ is closer to the estimates in the 
hidden and open charm and bottom hadrons~\cite{Ali:2017wsf}. This can be 
exemplified~\cite{Maiani:2015vwa} by the mass difference in hyperons 
$M_{\Lambda (1405)} - M_{\Lambda (1116)} \simeq 290$~MeV 
and in the exotic $X,\, Y,\, Z$ states where $\Delta M (L = 0 \to 1) \simeq 280$~MeV, 
 and similar differences in the charm baryons. 
As a conservative estimate of the uncertainty in~$B$, we took it as $\pm 20$~MeV, 
which is approximately the same as the uncertainty on the orbital coupling 
for the $\Omega_c^*$-baryons~\cite{Ali:2019xyz}.  
With these parameters, the mass of the third pentaquark $M = (4240 \pm 29)$~MeV 
with $J^P = 3/2^-$, also shown in the solid box, is somewhat lower than the mass 
of the observed $P_c (4312)^+$ peak, but is still in the right ball-park. 
A correlated error analysis on the pentaquark masses will be undertaken once 
the spin-parity of the three observed states is determined.

The second possibility is to assign the lowest mass state $P_c (4312)^+$ with the 
one from the orbitally-excited set of states, in particular, with the one having the 
spin-parity $J^P = 3/2^+$ and mass $M = 4360$~MeV, or with $J^P = 1/2^+$ and $M = 4351$~MeV. 
Both predictions are rather close to the observed pentaquark mass. However, in this case, some 
of the input parameters have unphysical values, and we do not entertain this assignment here.

The threshold for the observed pentaquarks in the $P_c^+ \to J/\psi + p$ decay mode 
is $M^{\rm thr}_{J/\psi\, p} = M_{J/\psi} + m_p = 4035.17$~MeV~\cite{Tanabashi:2018oca}. 
With the masses given in Table~\ref{tab:masses-predictions-cbar-cq-qq}, 
there are two states, the $J^P = 1/2^-$ with a mass  $3830$~MeV, and $J^P = 1/2^+$ 
with the mass $4031$~MeV, which lie below the $M^{\rm thr}_{J/\psi\, p}$ threshold.  
Also, a third state having $J^P=3/2^+$, with a mass 4040 MeV, may also lie below $M^{\rm thr}_{J/\psi\, p}$.
One of these states is even below the threshold for the decay $P_c^+ \to \eta_c + p$. 
Hence, it will decay weakly. The others shown in Table~\ref{tab:masses-predictions-cbar-cq-qq} 
are reachable in the $\Lambda_b \to J/\psi\, p\, K^-$ decay, and we urge the LHCb 
Collaboration to search for them.

\begin{table}[tb] 
\caption{
Masses of the hidden-charm unflavored pentaquarks (in MeV) and their 
comparison with the results presented in~\cite{Ali:2016dkf,Ali:2017ebb}. 
For the $P$-wave pentaquarks, following values of the parameters are used: 
orbital coupling $B = 207 \pm 20$~MeV and spin-orbit coupling of the heavy triquark 
$A_t = 3.4 \pm 1.1$~MeV. 
}
\label{tab:masses-predictions-cbar-cq-qq}
\begin{center}
\begin{tabular}{ccc} 
\hline
$J^P$ & \; This work \; & \; Refs.~\cite{Ali:2016dkf,Ali:2017ebb} \; \\ 
\hline 
\multicolumn{3}{c}{$S_{ld} = 0$, $L = 0$}  \\
$1/2^-$ &            $3830 \pm 34$ & $4086 \pm 42$  \\  
        &            $4150 \pm 29$ & $4162 \pm 38$  \\  
$3/2^-$ & \framebox{$4240 \pm 29$} & $4133 \pm 55$  \\ \cline{1-3} 
\multicolumn{3}{c}{$S_{ld} = 0$, $L = 1$}  \\
$1/2^+$ &            $4030 \pm 39$ & $4030 \pm 62$  \\  
        &            $4351 \pm 35$ & $4141 \pm 44$  \\  
        &            $4430 \pm 35$ & $4217 \pm 40$  \\  
$3/2^+$ &            $4040 \pm 39$ &                \\  
        &            $4361 \pm 35$ &                \\  
        & \framebox{$4440 \pm 35$} &                \\  
$5/2^+$ & \framebox{$4457 \pm 35$} & $4510 \pm 57$  \\ \hline 
\end{tabular}
\end{center}
\end{table} 

\section{Pentaquark Decay Widths} 
\label{sec:pentaquark-widths}  
In the compact diquark picture, the quarks in a diquark are bound and not free. 
In the present context it means that there is a barrier (or bound-state effect) which reduces the probability of the $\bar c$ and
the charm quark in the $[uc]$-diquark to form a charmonium state. This is seen also in the decays of the $X, Y, Z$ states, which
are tetraquark candidates in the compact diquark picture. The tunneling probability depends on the mass of the quark, with the
probabilty exponentially suppressed the heavier the quark is, expressed through the semi-classical approximation of the tunneling
amplitude $ A_M \sim e^{-\sqrt{2 M E l}}$, where $M$ is the quark mass, $l$ is the radius of a tetraquark (or pentaquark), of order
($1 -2 $) fm, and $E$ is typically 100 MeV~\cite{Maiani:2017kyi}. This suppresses the formation of the charmonium states, leading to a much reduced decay width for $P_c \to J/\psi p$, and less so for the decay $P_c \to \Lambda_c \bar D^{(*)}$,  which remains to be tested.
 In all these cases, the size of the diquark plays an important role 
in reducing the decay widths. For example, it is argued in ~\cite{Maiani:2017kyi}, that the relative ratio $\lambda$ between the
radius  of a tetraquark $R_{4q}$ and that of a compact diquark $R_{Qq}$ is expected to be $\lambda \equiv R_{4q}/R_{Qq} \geq 3$.
This presumably can also be taken as an estimate of this quantity for the pentaquarks. This will be checked also in the photoproduction experiments under way at the Jafferson Laboratory. The role of the photoproduction process $\gamma\, p \to J/\psi\, p$ in constraining the models of the new $P_c$-states is noted in~\cite{Cao:2019kst}.
We do expect reduced cross-sections for the narrow $P_c$-states in the compact diquark picture, as opposed to the one based 
on hadron molecule interpretation~\cite{Kubarovsky:2015aaa}.  Moreover,
as  some of the states are $P$-states, anticipated in the compact diquark picture, the decay widths 
are further reduced due to the angular momentum barrier.

\section{Conclusions} 
\label{sec:conclusions} 

We have presented the mass spectrum of the hidden-charm  pentaquark states 
$(c \bar c q q^\prime q^{\prime\prime})$, where~$q$, $q^\prime$, and~$q^{\prime\prime}$ 
are light $u$- and $d$-quarks, using the isospin symmetry. In doing this, we have used
an effective Hamiltonian, based on a doubly-heavy triquark~--- light diquark picture 
of a pentaquark, shown in Fig.~\ref{ali:fig-pentaquark-model-2}. Apart from the constituent
quark and diquark masses, the Hamiltonian incorporates dominant spin-spin, spin-orbit 
and orbital interactions. 
Following this, we interpret the three narrow resonances with the states having the following 
spin and angular momentum quantum numbers: 
$P_c (4312)^+ = \{\bar c [cu]_{s=1} [ud]_{s=0}; L_{\mathcal{P}} = 0, J^P = 3/2^- \}$, the $S$-wave,
and the other two as $P$-wave states, with 
$P_c (4440)^+ = \{\bar c [cu]_{s=1} [ud]_{s=0}; L_{\mathcal{P}} = 1, J^P = 3/2^+ \}$, and
$P_c (4457)^+ = \{\bar c [cu]_{s=1} [ud]_{s=0}; L_{\mathcal{P}} = 1, J^P = 5/2^+ \}$.
We have presented analytical expressions for the masses of the states having well-defined quantum
numbers.
 However, there are more states present in the spectrum having 
the quark content $(c \bar c u u d)$. They are listed in Tables~\ref{tab:S-wave-pentaquarks-good-ld} 
and~\ref{tab:P-wave-pentaquarks-good-ld}. The details of the formalism are given 
elsewhere~\cite{Ali:2019xyz}, where also the spectrum of the hidden-charm pentaquarks, having 
a non-vanishing strangeness  
is worked out, as well as the states having spin-1 light diquarks, i.\,e., with $S_{ld} = 1$. 
Some of them can be reached in weak decays of bottom-strange baryons,~$\Xi_b$ and~$\Omega_b$~\cite{Ali:2016dkf}.

For the numerical estimates of the pentaquark masses, we infer the input parametric values 
from the earlier studies of the hidden-charm tetraquarks and charmed baryons. Since not all 
parameters in the effective Hamiltonian can be uniquely determined from the existing resonances, 
as there are currently only three such observed states, we use additionally the results of 
the known heavy baryons, such as the orbitally-excited $\Omega_c^*$ states, since $P_c (4312)^+$, 
$P_c (4440)^+$ and $P_c (4457)^+$ are also heavy baryons. It is difficult to quantify the 
theoretical errors on our mass estimates, as we have dropped sub-dominant spin-spin interactions, 
but we trust that the masses of the pentaquarks given in Table~\ref{tab:masses-predictions-cbar-cq-qq} 
are in the right ball-park. We note that, in addition to the observed ones, there are three states 
(one with $J^P = 1/2^-$ and two with positive parity, $J^P = 1/2^+$ and $J^P = 3/2^+$) which have $[cu]_{s=0}$-diquark in their Fock space. 
Their masses are lower than their counterparts with $[cu]_{s=1}$-diquark. Two of these states lie 
below the $J/\psi\, p$ threshold, probably the third one as well, requiring a different search strategy. We anticipate other 
pentaquark states waiting to be discovered with yet more data and a detailed study of the 
$J/\psi\, p$ mass spectrum, including both narrow and broad resonances.

Finally, we stress that it is not enough to claim the observed narrow peaks in the $J/\psi\, p$ 
mass spectrum as a confirmation of hadronic molecules, just due to their kinematic vicinity
to the charmed meson-charmed baryon thresholds, $\Sigma_c\, \bar{D}$ and $\Sigma_c\, \bar{D}^*$, 
which in this hypothesis are interpreted as loosely-bound $S$-wave states, and hence have 
necessarily a negative parity. Crucially, the spin-parities of the $P_c (4312)^+$, $P_c (4440)^+$ 
and $P_c (4457)^+$ remain to be determined. In the compact pentaquark picture presented here, 
heavy-quark symmetry requirements are built in and the mass estimates are in the right ball-park. 
The broad resonance $P_c(4380)^+$, present in the older LHCb analysis~\cite{Aaij:2015tga}, 
which was difficult to accommodate in the compact pentaquark picture due to the conflict 
with the heavy quark symmetry~\cite{Ali:2016dkf}, is presumably no longer there. So, the compact pentaquark
interpretation of the three $P_c$-resonances is very much in the game. In particular, and as 
a crucial test, we anticipate the three narrow resonances to have different parities, i.\,e., 
both negative and positive parities are foreseen. Their determination would help greatly 
in discriminating the compact pentaquark interpretation and the competing ones based on hadronic 
molecules in various incarnations~\cite{Chen:2019bip,Chen:2019asm,Guo:2019fdo,Liu:2019tjn,He:2019ify,Xiao:2019aya,%
Huang:2019jlf,Shimizu:2019ptd,Xiao:2019mvs}.

\bigskip

\textbf{Acknowledgments}

We thank Luciano Maiani, Antonello Polosa and Sheldon Stone 
for reading the manuscript and helpful discussions.
A.\,P. thanks the Theory Group at DESY for their kind hospitality, where the major 
part of this work was done, and acknowledges financial support by the German Academic 
Exchange Service (DAAD), by the Russian Foundation for Basic Research and National 
Natural Science Foundation of China for the research project No. 19-52-53041, and by the ``YSU Initiative 
Scientific Research Activity'' (Project No. AAAA-A16-116070610023-3).


\begin{thebibliography}{99}
%


\bibitem{Aaij:2015tga} 
  R.~Aaij {\it et al.} [LHCb Collaboration],
  Phys.\ Rev.\ Lett.\  {\bf 115}, 072001 (2015)
  [arXiv:1507.03414 [hep-ex]].

\bibitem{Aaij:2016ymb} 
  R.~Aaij {\it et al.} [LHCb Collaboration],
  Phys.\ Rev.\ Lett.\  {\bf 117}, 082003 (2016)
  Addendum: [Phys.\ Rev.\ Lett.\  {\bf 117}, 109902 (2016)]
  Addendum: [Phys.\ Rev.\ Lett.\  {\bf 118}, 119901 (2017)]
  [arXiv:1606.06999 [hep-ex]].

\bibitem{Tanabashi:2018oca} 
  M.~Tanabashi {\it et al.} [Particle Data Group],
  Phys.\ Rev.\ D {\bf 98}, 030001 (2018).
  
\bibitem{Guo:2015umn} 
  F.\,K.~Guo, U.\,G.~Meissner, W.~Wang and Z.~Yang,
  Phys.\ Rev.\ D {\bf 92}, 071502 (2015)
  [arXiv:1507.04950 [hep-ph]].
%
\bibitem{Liu:2015fea} 
  X.\,H.~Liu, Q.~Wang and Q.~Zhao,
  Phys.\ Lett.\ B {\bf 757}, 231 (2016)
  [arXiv:1507.05359 [hep-ph]].
%
\bibitem{Mikhasenko:2015vca} 
  M.~Mikhasenko,
  arXiv:1507.06552 [hep-ph].
%
\bibitem{Meissner:2015mza} 
  U.\,G.~Meissner and J.\,A.~Oller,
  Phys.\ Lett.\ B {\bf 751}, 59 (2015)
  [arXiv:1507.07478 [hep-ph]].

%
\bibitem{Chen:2015moa} 
  H.\,X.~Chen, W.~Chen, X.~Liu, T.\,G.~Steele and S.\,L.~Zhu,
  Phys.\ Rev.\ Lett.\  {\bf 115}, 172001 (2015)
  [arXiv:1507.03717 [hep-ph]].
%
\bibitem{He:2015cea} 
  J.~He,
  Phys.\ Lett.\ B {\bf 753}, 547 (2016)
  [arXiv:1507.05200 [hep-ph]].
%
\bibitem{Roca:2015dva} 
  L.~Roca, J.~Nieves and E.~Oset,
  Phys.\ Rev.\ D {\bf 92}, 094003 (2015)
  [arXiv:1507.04249 [hep-ph]].
%
\bibitem{Chen:2015loa} 
  R.~Chen, X.~Liu, X.\,Q.~Li and S.\,L.~Zhu,
  Phys.\ Rev.\ Lett.\  {\bf 115}, 132002 (2015)
  [arXiv:1507.03704 [hep-ph]].
%
\bibitem{Xiao:2015fia} 
  C.\,W.~Xiao and U.-G.~Meissner,
  Phys.\ Rev.\ D {\bf 92}, 114002 (2015)
  [arXiv:1508.00924 [hep-ph]].

\bibitem{Kubarovsky:2015aaa} 
  V.~Kubarovsky and M.\,B.~Voloshin,
  Phys.\ Rev.\ D {\bf 92}, 031502 (2015)
  [arXiv:1508.00888 [hep-ph]].
  
\bibitem{Li:2015gta} 
  G.\,N.~Li, X.\,G.~He and M.~He,
  JHEP {\bf 1512}, 128 (2015)
  [arXiv:1507.08252 [hep-ph]].
%
\bibitem{Mironov:2015ica} 
  A.~Mironov and A.~Morozov,
  JETP Lett.\  {\bf 102}, 271 (2015)
  [Pisma Zh.\ Eksp.\ Teor.\ Fiz.\  {\bf 102}, 302 (2015)]
  [arXiv:1507.04694 [hep-ph]].
%
\bibitem{Anisovich:2015cia} 
  V.\,V.~Anisovich, M.\,A.~Matveev, J.~Nyiri, A.\,V.~Sarantsev and A.\,N.~Semenova,
  arXiv:1507.07652 [hep-ph].
%
\bibitem{Ghosh:2015ksa} 
  R.~Ghosh, A.~Bhattacharya and B.~Chakrabarti,
  Phys.\ Part.\ Nucl.\ Lett.\  {\bf 14}, 550 (2017)
  [arXiv:1508.00356 [hep-ph]].
%
\bibitem{Wang:2015epa} 
  Z.\,G.~Wang,
  Eur.\ Phys.\ J.\ C {\bf 76}, 70 (2016)
  [arXiv:1508.01468 [hep-ph]].
%
\bibitem{Wang:2015ava} 
  Z.\,G.~Wang and T.~Huang,
  Eur.\ Phys.\ J.\ C {\bf 76}, 43 (2016)
  [arXiv:1508.04189 [hep-ph]].
%
\bibitem{Wang:2015wsa} 
  Z.\,G.~Wang,
  Eur.\ Phys.\ J.\ C {\bf 76}, 142 (2016)
  [arXiv:1509.06436 [hep-ph]].
%

\bibitem{Maiani:2015vwa} 
  L.~Maiani, A.\,D.~Polosa and V.~Riquer,
  Phys.\ Lett.\ B {\bf 749}, 289 (2015)
  [arXiv:1507.04980 [hep-ph]].
  
\bibitem{Ali:2016dkf} 
  A.~Ali, I.~Ahmed, M.\,J.~Aslam and A.~Rehman,
  Phys.\ Rev.\ D {\bf 94}, 054001 (2016)
  [arXiv:1607.00987 [hep-ph]].
  
\bibitem{Ali:2017ebb} 
  A.~Ali, I.~Ahmed, M.\,J.~Aslam and A.~Rehman,
  arXiv:1704.05419 [hep-ph].
  
\bibitem{Lebed:2015tna} 
  R.\,F.~Lebed,
  Phys.\ Lett.\ B {\bf 749}, 454 (2015)
  [arXiv:1507.05867 [hep-ph]].
 
\bibitem{Zhu:2015bba} 
  R.~Zhu and C.\,F.~Qiao,
  Phys.\ Lett.\ B {\bf 756}, 259 (2016)
  [arXiv:1510.08693 [hep-ph]].


\bibitem{Aaij:2019vzc} 
  R.~Aaij {\it et al.} [LHCb Collaboration],
  arXiv:1904.03947 [hep-ex].

  

\bibitem{Maiani:2004vq} 
  L.~Maiani, F.~Piccinini, A.\,D.~Polosa and V.~Riquer,
  Phys.\ Rev.\ D {\bf 71}, 014028 (2005)
  [hep-ph/0412098].
  
\bibitem{Lipkin:1987sk} 
  H.\,J.~Lipkin,
  Phys.\ Lett.\ B {\bf 195}, 484 (1987).

\bibitem{Jaffe:2003sg} 
  R.\,L.~Jaffe and F.~Wilczek,
  Phys.\ Rev.\ Lett.\  {\bf 91}, 232003 (2003)
  [hep-ph/0307341].
  
\bibitem{Eichten:1978tg} 
  E.~Eichten, K.~Gottfried, T.~Kinoshita, K.\,D.~Lane and T.\,M.~Yan,
  Phys.\ Rev.\ D {\bf 17}, 3090 (1978)
  Erratum: [Phys.\ Rev.\ D {\bf 21}, 313 (1980)].

\bibitem{Manohar:1992nd} 
  A.\,V.~Manohar and M.\,B.~Wise,
  Nucl.\ Phys.\ B {\bf 399}, 17 (1993)
  [hep-ph/9212236].
  
\bibitem{Esposito:2013fma} 
  A.~Esposito, M.~Papinutto, A.~Pilloni, A.\,D.~Polosa and N.~Tantalo,
  Phys.\ Rev.\ D {\bf 88}, 054029 (2013)
  [arXiv:1307.2873 [hep-ph]].


\bibitem{Luo:2017eub} 
  S.\,Q.~Luo, K.~Chen, X.~Liu, Y.\,R.~Liu and S.\,L.~Zhu,
  Eur.\ Phys.\ J.\ C {\bf 77}, 709 (2017)
  [arXiv:1707.01180 [hep-ph]].

\bibitem{Karliner:2017qjm} 
  M.~Karliner and J.\,L.~Rosner,
  Phys.\ Rev.\ Lett.\  {\bf 119}, 202001 (2017)
  [arXiv:1707.07666 [hep-ph]].

\bibitem{Eichten:2017ffp} 
  E.\,J.~Eichten and C.~Quigg,
  Phys.\ Rev.\ Lett.\  {\bf 119}, 202002 (2017)
  [arXiv:1707.09575 [hep-ph]].

\bibitem{Francis:2016hui} 
  A.~Francis, R.\,J.~Hudspith, R.~Lewis and K.~Maltman,
  Phys.\ Rev.\ Lett.\  {\bf 118}, 142001 (2017)
  [arXiv:1607.05214 [hep-lat]].


\bibitem{Bicudo:2017szl} 
  P.~Bicudo, M.~Cardoso, A.~Peters, M.~Pflaumer and M.~Wagner,
  Phys.\ Rev.\ D {\bf 96}, 054510 (2017)
  [arXiv:1704.02383 [hep-lat]].

\bibitem{Junnarkar:2017sey} 
  P.~Junnarkar, M.~Padmanath and N.~Mathur,
  EPJ Web Conf.\  {\bf 175}, 05014 (2018)
  [arXiv:1712.08400 [hep-lat]].
  
\bibitem{Mehen:2017nrh} 
  T.~Mehen,
  Phys.\ Rev.\ D {\bf 96}, 094028 (2017)
  [arXiv:1708.05020 [hep-ph]].


\bibitem{Czarnecki:2017vco} 
  A.~Czarnecki, B.~Leng and M.\,B.~Voloshin,
  Phys.\ Lett.\ B {\bf 778}, 233 (2018)
  [arXiv:1708.04594 [hep-ph]].

\bibitem{Maiani:2019cwl} 
  L.~Maiani, A.\,D.~Polosa and V.~Riquer,
  arXiv:1903.10253 [hep-ph].


\bibitem{Maiani:2014aja} 
  L.~Maiani, F.~Piccinini, A.\,D.~Polosa and V.~Riquer,
  Phys.\ Rev.\ D {\bf 89}, 114010 (2014)
  [arXiv:1405.1551 [hep-ph]].

\bibitem{Ali:2017wsf} 
  A.~Ali, L.~Maiani, A.\,V.~Borisov, I.~Ahmed, M.\,J.~Aslam, A.\,Ya.~Parkhomenko,
  A.\,D.~Polosa and A.~Rehman,
  Eur.\ Phys.~J.~C {\bf 78}, 29 (2018)
  arXiv:1708.04650 [hep-ph].

\bibitem{Ali:2019xyz} 
  A.~Ali, I.~Ahmed, M.\,J.~Aslam, A.\,Ya.~Parkhomenko, and A.~Rehman,
  DESY 19-052 (April 2019).
  
\bibitem{Ali:2019}
  A.~Ali, L.~Maiani, and A.\,D.~Polosa,
  {\it Multiquark Hadrons}.   
  Cambridge University Press, Cambridge, 2019. 
  
\bibitem{Karliner:2014gca} 
  M.~Karliner and J.\,L.~Rosner,
  Phys.\ Rev.\ D {\bf 90}, 094007 (2014)
  [arXiv:1408.5877 [hep-ph]].
  
\bibitem{Karliner:2017kfm} 
  M.~Karliner and J.\,L.~Rosner,
  Phys.\ Rev.\ D {\bf 95}, 114012 (2017).
  [arXiv:1703.07774 [hep-ph]].
 

\bibitem{Karliner:2018bms} 
  M.~Karliner and J.\,L.~Rosner,
  Phys.\ Rev.\ D {\bf 98}, 074026 (2018)
  [arXiv:1808.07869 [hep-ph]].

\bibitem{Maiani:2017kyi} 
  L.~Maiani, A.~D.~Polosa and V.~Riquer,
  Phys.\ Lett.\ B {\bf 778}, 247 (2018)
  [arXiv:1712.05296 [hep-ph]].

\bibitem{Cao:2019kst}
  X.~Cao and J.\,p.~Dai,
  arXiv:1904.06015 [hep-ph].
\bibitem{Chen:2019bip} 
  H.\,X.~Chen, W.~Chen and S.\,L.~Zhu,
  arXiv:1903.11001 [hep-ph].

\bibitem{Chen:2019asm} 
  R.~Chen, X.~Liu, Z.\,F.~Sun and S.\,L.~Zhu,
  arXiv:1903.11013 [hep-ph].
 
\bibitem{Guo:2019fdo} 
  F.\,K.~Guo, H.\,J.~Jing, U.\,G.~Meissner and S.~Sakai,
  arXiv:1903.11503 [hep-ph].
 
\bibitem{Liu:2019tjn} 
  M.\,Z.~Liu, Y.\,W.~Pan, F.\,Z.~Peng, M.~Sanchez Sanchez, L.\,S.~Geng, A.~Hosaka and M.\,P.~Valderrama,
  arXiv:1903.11560 [hep-ph].
  
\bibitem{He:2019ify} 
  J.~He,
  arXiv:1903.11872 [hep-ph].

\bibitem{Xiao:2019aya} 
  C.\,W.~Xiao, J.~Nieves and E.~Oset,
  arXiv:1904.01296 [hep-ph].

\bibitem{Huang:2019jlf} 
  H.~Huang, J.~He and J.~Ping,
  arXiv:1904.00221 [hep-ph].
 
 
\bibitem{Shimizu:2019ptd} 
  Y.~Shimizu, Y.~Yamaguchi and M.~Harada,
  arXiv:1904.00587 [hep-ph].

\bibitem{Xiao:2019mvs} 
  C.\,J.~Xiao, Y.~Huang, Y.\,B.~Dong, L.\,S.~Geng and D.\,Y.~Chen,
  arXiv:1904.00872 [hep-ph].


\end{thebibliography}
\end{document}